\documentclass[iop,apj]{emulateapj}
\shorttitle{TURBULENCE IN THE SUPERMODEL}
\shortauthors{R. FUSCO-FEMIANO \& A. LAPI}

\begin{document}

\title{Turbulence in the SuperModel:\\ Mass Reconstruction with Nonthermal Pressure for Abell 1835}
\author{R. Fusco-Femiano$^{1}$ and A. Lapi$^{2,3}$}
\affil{$^1$IAPS-INAF, Via Fosso del Cavaliere, 00133 Roma,
Italy.\\$^2$Dip. Fisica, Univ. `Tor Vergata', Via Ricerca Scientifica
1, 00133 Roma, Italy.\\$^3$SISSA, Via Bonomea 265, 34136 Trieste, Italy.}

\begin{abstract}
The total mass derived from X-ray emission is biased low in a large number
of clusters when compared with the mass estimated via strong and weak
lensing. \textsl{Suzaku} and \textsl{Chandra} observations out to the virial radius report
in several relaxed clusters steep temperature gradients that on assuming
pure thermal hydrostatic equilibrium imply an unphysically decreasing mass
profile. Moreover, the gas mass fraction appears to be inconsistent with
the cosmic value measured from the CMB. Such findings can be interpreted as
an evidence for an additional nonthermal pressure in the outskirts of these
clusters. This nonthermal component may be due to turbulence stirred by
residual bulk motions of extragalactic gas infalling into the cluster. Here
we present a SuperModel analysis of Abell 1835 observed by \textsl{Chandra} out to
the virial radius. The SuperModel formalism can include in the equilibrium
a nonthermal component whose level and distribution are derived imposing
that the gas mass fraction $f_{\rm gas}$ equals the cosmic value at the
virial radius. Including such a nonthermal component, we reconstruct from X
rays an increasing mass profile consistent with the hydrostatic equilibrium
also in the cluster outskirts and in agreement at the virial boundary with
the weak lensing value. The increasing $f_{\rm gas}$ profile confirms that the
baryons are not missing but located at the cluster outskirts.
\end{abstract}

\keywords{cosmic background radiation --- galaxies: clusters: individual
(Abell 1835) --- X-rays: galaxies: clusters}

\section{Introduction}

Clusters of galaxies formed from the collapse of primordial density
fluctuations are powerful cosmological probes mostly relied on the their
total virial mass. The traditional method to estimate $M(r)$ is based on the
IntraCluster Plasma (ICP) density and temperature profiles derived from the
X-ray bremsstrahlung emission. These profiles allow to solve the equation of
hydrostatic equilibrium (HE) assuming spherical symmetry. The comparison with
masses estimated via strong and weak lensing (Arnaud et al. 2007; Mahdavi
et al. 2008; Lau et al. 2009; Battaglia et al. 2012) has highlighted
that the X-ray mass is biased low by a systematic $\sim$ 10-20$\%$ even in
relaxed clusters. These differences in the mass values suggest the presence
of a nonthermal gas pressure support that could resolve this discrepancy. On
the other hand, simulations unanimously show the presence of gas motions
driven by inflow of material into the cluster from its environment, by
mergers, and by the supersonic movements of galaxies through the ICP. These
motions may cause the development of turbulence in the cluster outskirts with
a deep impact on the physics of the ICP (Nagai et al. 2007; Shaw et al. 2010;
Burns et al. 2010; Vazza et al. 2011; Rasia et al. 2012). Also the gas clumping,
that may be important at large radii, can considerably underestimate the
total mass (Nagai \& Lau 2011; Simionescu et al. 2011; Eckert et al. 2012;
Vazza et al. 2013).

An incorrect estimate of the total mass implies an incorrect determination of
the baryon fraction $f_{\rm gas}$ in the ICP that contains most of the baryons
in clusters. The remaning baryons with a fraction $f_{\rm stars}$ that represent a few percent
of the total mass are in stars and intracluster light (Gonzalez et al. 2007;
Giodini et al. 2009). The total baryon fraction ($f_b = f_{\rm gas} + f_{\rm stars}$)
and its evolution with the red-shift are used to constrain cosmological
parameters since it is believed to be representative of the Universe (e.g.,
White et al. 1993; Metzler \& Evrard 1994; Ettori et al. 2009). Current studies
have shown that the cluster baryon fraction $f_b$ derived at $r_{500}$ is
lower than the ratio $\Omega_b/\Omega_M$ measured from the CMB by several
experiments (Afshordi et al. 2007; Umetsu et al. 2009; Vikhlinin et al. 2006;
Arnaud et al. 2007; Sun et al. 2009; Komatsu et al. 2011) raising the question
of where the missing baryons are allocated (Rasheed et al. 2010). To address
this issue, Landry et al. (2012) have recently used \textsl{Chandra} X-ray
observations to measure the gas mass fraction for a complete sample of
massive clusters in the red-shift range (0.15- 0.30) from the
\textit{Brightest Cluster Sample} (Ebeling et al. 1998; Dahle 2006). These
clusters are observed at the radius within which the mass density is 500
times the critical density of the universe at the cluster's redshift. They
find that the baryon content in these high-luminosity clusters is consistent
at $r_{500}$ with the cosmic ratio $\Omega_b/\Omega_M = 0.167 \pm 0.006$
implying that there are no missing baryons within this radius in the most
luminous and massive clusters. But, in accord with several studies they
measure an increase of $f_{\rm gas}$ with radius raising the question of what
happens to the gas mass fraction beyond $r_{500}$. It can be presumed always
higher values of $f_{\rm gas}$ going toward the virial boundary as also reported
by recent \textsl{Suzaku} observations (e.g., Simionescu et al. 2011). However, Landry
et al. (2012) doubt the validity of this extrapolation considering that the
gas could not be in HE beyond $r_{500}$, and/or that the clumping of the gas
may become always more important toward the virial radius. An underestimate
of the total mass may be the cause of the discrepancy between $f_b$ and the
ratio $\Omega_b/\Omega_M$ at $r > r_{500}$.

One of the clusters in the sample of Landry et al. (2012) is Abell 1835 ($z$ =
0.253) that has been investigated by Bonamente et al. (2013) out to the virial
radius thanks to a long exposure and high quality data. \textsl{Chandra} reports soft
X-ray surface brightness emission out to a radial distance of $\sim$ 2.4 Mpc,
and a very step temperature profile like to that observed by \textsl{Suzaku} in some
relaxed clusters (Bautz et al. 2009; Hoshino et al. 2010; Kawarada et al. 2010;
Walker et al. 2012). This temperature profile implies a decreasing total mass
profile at $r > r_{500}$, a $f_b$ value consistent with the cosmological
ratio at $r_{500}$, but inconsistent at greater distances. Their conclusion
is that the steepening of the temperature profile is incompatible with the HE
in the outskirts of the cluster as confirmed by recent \textsl{Suzaku} observations
out to the virial radius (Ichikawa et al. 2013). Besides, Bonamente et al.
(2013), report that a negative entropy gradient renders the ICP convectively
unstable, flattening within a few Gyrs the temperature profile for the
transport of central hotter gas in the periphery of the cluster. They suggest
the presence of cool gas in the outskirts of the cluster that may be the
result of infall from the filamentary structure if this gas lies in
projection against the outermost regions.

Here we show how it is possible to reconstruct the total cluster mass using
the SuperModel (Cavaliere et al. 2009) that includes a nonthermal pressure
component (Cavaliere et al. 2011) due to turbulent motions. This component in
addition to the ICP thermal pressure sustains the HE. Turbulence is related
to the weakening of the accretion shocks that induces an increase of the bulk
inflow energy in the cluster outskirts and also saturation of entropy
production determining the observed steep temperature profiles (see Lapi
et al. 2010). In particular, we analyze Abell 1835 showing that the inclusion
of this nonthermal component gives an increasing total mass also in the
cluster outskirts and in agreement with the weak lensing measurements. The
level and distribution of this nonthermal pressure support are obtained
imposing that the baryon mass fraction is consistent with the cosmic ratio at
the virial boundary (see Sect. 3).

Throughout the paper we adopt the standard flat cosmology with parameters
$H_0 = 70$ km s$^{-1}$ Mpc$^{-1}$, $\Omega_{\Lambda} = 0.7$, $\Omega_M = 0.3$
(see Komatsu et al. 2011; Hinshaw et al. 2013, \textsl{Planck} collaboration 2013a). With
this 1 arcmin corresponds to 237.48 kpc.

\section{Turbulence in the SuperModel}

The wealth of current and upcoming data for emission in X-rays and scattering
in the $\mu$waves of the CMB photons for the Sunyaev-Zel$^{,}$dovich effect
requires a precision modelling of the ICP density $n(r)$ and temperature
$T(r)$ distributions. This modeling is provided by the SM based on the run of
the ICP specific \textit{entropy} (adiabat) $k = k_B T/n^{2/3}$ set by the
processes for its production and erosion. AGN outbursts and deep mergers
often followed by inner sloshing determine a raise of the entropy at the
cluster centers; in addition there the entropy may be partly eroded by
cooling processes. At the other end, a large quantity of entropy is
continuously produced at the virial boundary $R$ where the ICP is shocked by
the supersonic gravitational inflow of gas accreted from the environment
along with the dark matter (DM), and is adiabatically stratified into the DM
potential well. These physical processes concur to create a spherically
averaged profile for the ICP entropy $k(r) = k_c + (k_R - k_c)(r/R)^a$, see
Voit (2005). A central floor $k_c$ ($\approx$ 10-100 keV cm$^2$) is followed
by an outer ramp with slope $a \approx$ 1 (Tozzi \& Norman 2001) leading to
entropy values $k_R \sim$ some 10$^3$ keV cm$^2$ at the virial boundary.

The thermal pressure $p(r) \propto k(r) n^{5/3}(r)$ is used in the SM to
balance the DM gravitational pull $-GM(< r)/r^2$ and sustain the hydrostatic
equilibrium out to the virial boundary. From the HE equation we directly
derive the temperature profile:
\begin{equation}
\frac{T(r)}{T_R} = \left[\frac{k(r)}{k_R}\right]^{3/5} \left\{ 1 + \frac{2}{5}
b_R\int_r^R {\frac{\mathrm{d}x}{x}~\frac{v^2_c(x)}{v^2_R}\,\left[\frac{k_R}{k(x)}\right]^{3/5}}\right\}~.
\end{equation}

Note that the density follows $n(r) = [k_BT(r)/k(r)]^{3/2}$, so that $T(r)$
and $n(r)$ are linked, rather than independently rendered with
multiparametric expressions as in other approaches. The few physical
parameters specifying $k(r)$ are enough to provide remarkably good fits to
the detailed X-ray data on surface brightness and on temperature profiles of
many cool-core CCs and non-cool-core NCCs clusters (see Fusco-Femiano et al.
2009), and to the SZ \textsl{Planck} profile for the Coma cluster (Fusco-Femiano et al.
2013). Good fits have been also obtained for the steep temperature profiles
observed by \textsl{Suzaku} out to the virial radius in some relaxed CC clusters
that, as suggested by Lapi et al. (2010), can be explained in terms of the
entropy profile flattening observed in these clusters. The entropy run starts
with an initial slope $a$, but for $r > r_b$ it deviate downward from a
simple powerlaw (see Eq.~4 in Lapi et al. 2010) where $r_b$ is a free
parameter. The outer branch of the entropy profile is described by a linear
decline of the slope with gradient $a^{\prime} \equiv (a-a_R)/(R/r_b - 1)$.
This lower entropy production may be explained in terms of decreasing
accretion rate due to the slowdown at later cosmic times of the cosmological
structure growth in an accelerating Universe. The effect is enhanced by
little mass available for accretion in cluster sectors adjacent to
low-density regions of the surrounding environment. So, we expect azimuthal
variations of the X-ray observables (Lapi et al. 2010).

This scenario seems confirmed by a recent analysis of a sample of relaxed
cool-core clusters at redshift below 0.25 (Walker et al. 2012). On the other
hand, the clumping effect reported by numerical simulations (Nagai \& Lau
2011) is not large enough to account for the observed amount of entropy
flattening. Also the proposed difference between the electron and ion
temperatures in the ICP inside the accretion shock in the outskirts as a
cause of the entropy profile flattening (Hoshino et al. 2010; Akamatsu et al.
2011) seems to be in contrast with Sunyaev-Zeldovich (SZ) effect observations
with \textsl{Planck} (\textsl{Planck} collaboration 2013b).

\begin{figure*}
\begin{center}
\epsscale{1}\plottwo{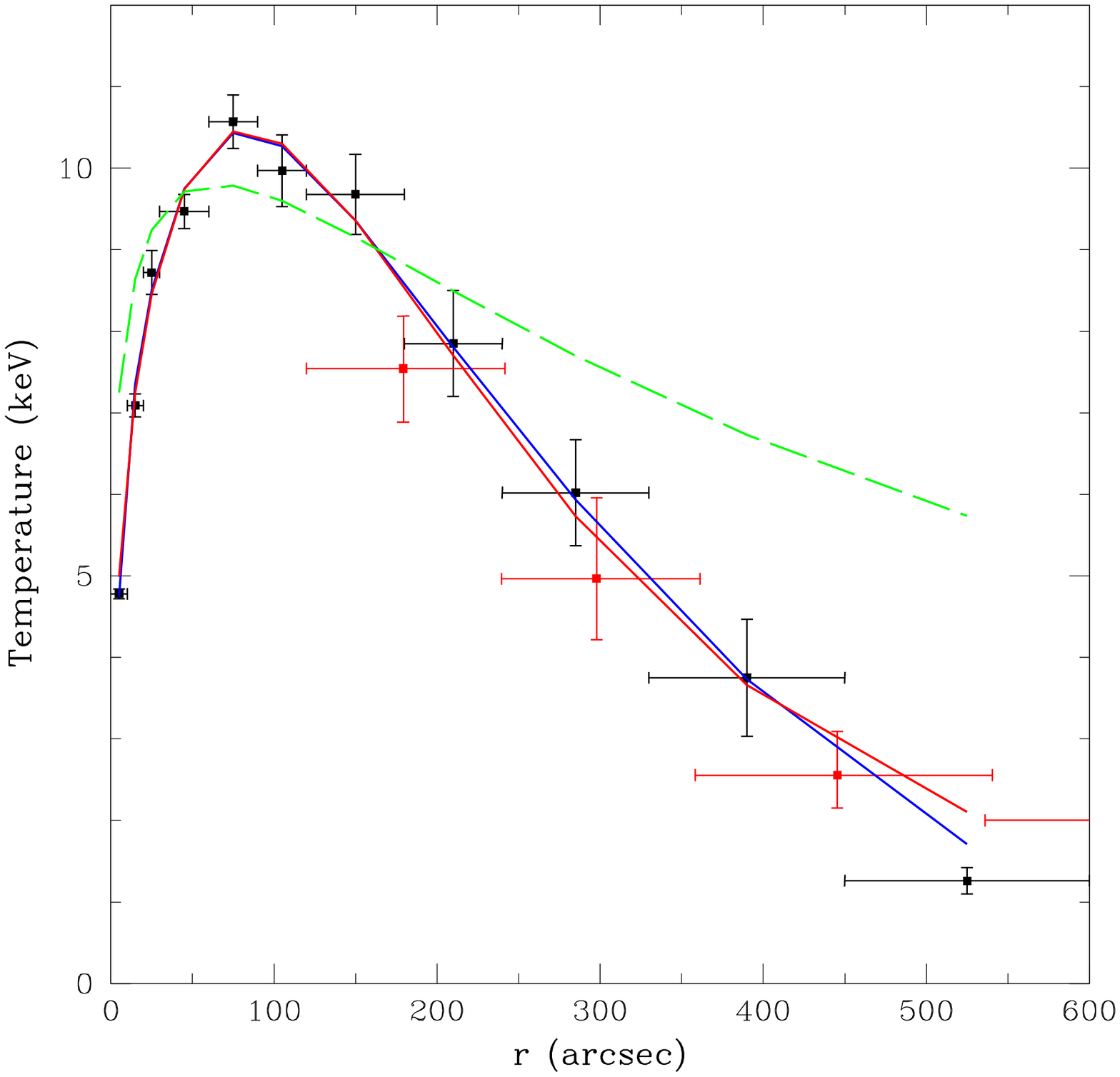}{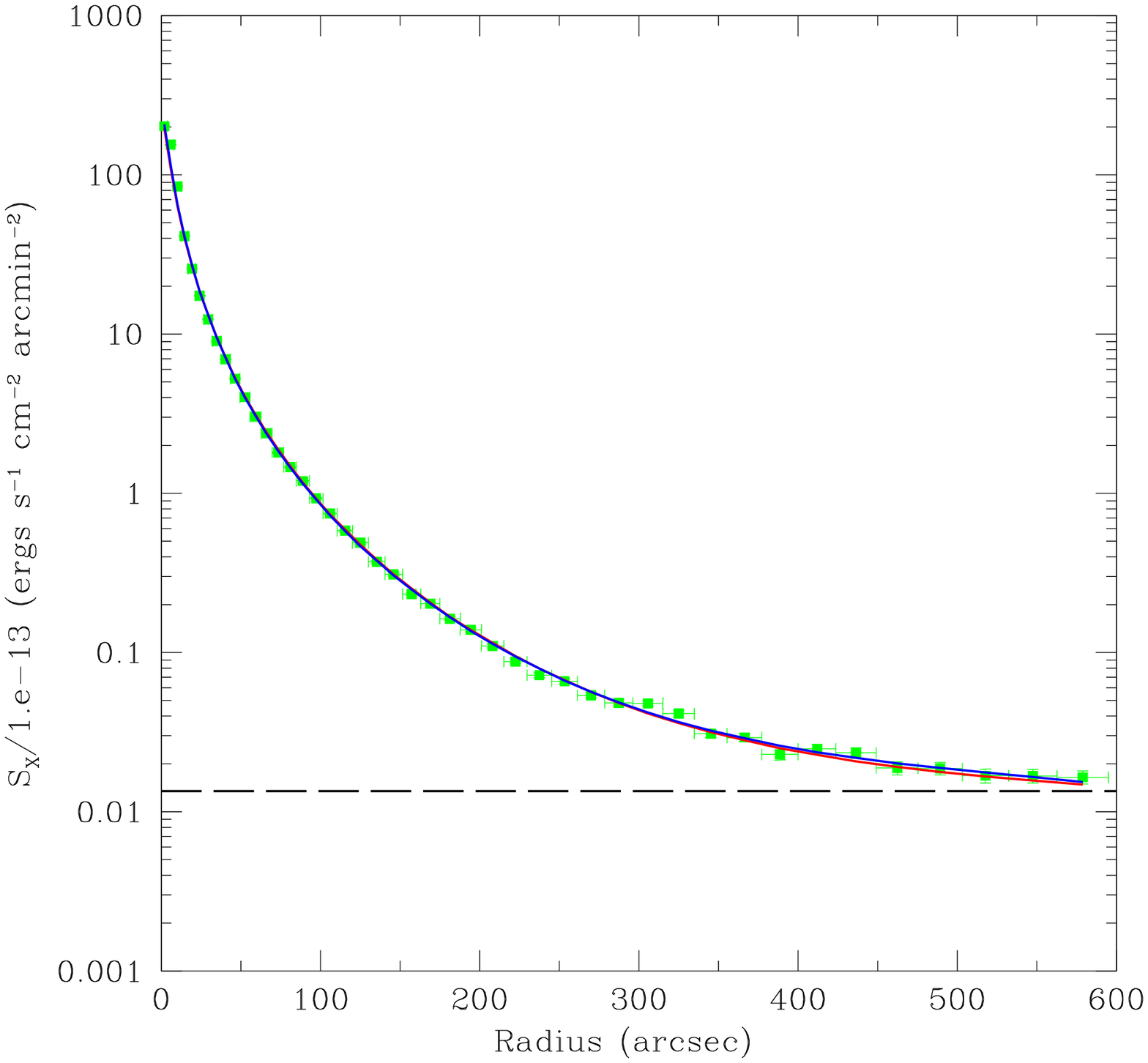}\caption{Left panel: Projected
temperature profile (black points) observed by \textsl{Chandra} in Abell 1835
(Bonamente et al. 2012); red points are by \textsl{Suzaku} (Ichikawa et al. 2013). Blue
line is the SM fit with $\delta(r)$ = 0 (see Eq.~1); red line is the SM fit
with $\delta_R$ = 1.4 and $\ell$ = 0.5 (see Eqs.~2 and 3); green line is the fit
without imposing the entropy flattening at $r > r_b$ (see text). Right
panel: Exposure corrected surface brightness profile of Abell 1835 in the
X-ray band (0.7-2 keV) observed by \textsl{Chandra}; the dashed line is the
background level (Bonamente et al. 2012); blue line is the SM fit with
$\delta(r)$ = 0 (see Eq.~1); red line is the SM fit with the above values of
$\delta_R$ and $\ell$.}
\end{center}
\end{figure*}

The weakening of the accretion shock in relaxed clusters not only
reduces the thermal energy to feed the intracluster entropy, but also
increases the amount of bulk energy to drive turbulence into the outskirts
(Cavaliere et al. 2011). Turbulent motions start a the virial radius $R$ with
coherence lenghts $L \sim R$/2 set in relaxed CC clusters by the pressure
scale height or by shock segmentation (see Iapichino \& Niemeyer 2008;
Valdarnini 2010; Vazza et al. 2010). Then they fragment downstream into a
dispersive cascade to sizes $\ell$. Larger values of turbulent energy
compared to the gas thermal energy are reported by simulations in the
innermost cluster regions of post-merger and merging clusters. Here instead
we deal with the outskirts of relaxed clusters where the simulations report
much lower values in the cluster cores but an increasing
$E_{\rm turb}/E_{\rm thermal}$ profile going toward the virial radius (e.g., Vazza
et al. 2011).

Since turbulent motions contribute to the pressure to substain HE, we focus
on the ratio $\delta(r) \equiv p_{\rm nth}/p_{\rm th}$ of turbulent to thermal
pressure with radial shape decaying on the scale $\ell$ from the boundary value
$\delta_R$. The total pressure is now $p_{\rm tot}(r) = p_{\rm th}(r) + p_{\rm nth}(r) =
p_{\rm th}(r)[1 + \delta(r)]$ that when inserted in the HE equation gives the
temperature profile in the form
\begin{eqnarray}
\nonumber\frac{T(r)}{T_R} &=& \left[\frac{k(r)}{k_R}\right]^{3/5}\,\left[\frac{1 + \delta_R}{1 +
\delta(r)}\right]^{2/5}\, \left\{1+\frac{2}{5}\,\frac{b_R}{1 + \delta_R}\times\right.\\
&&\\
\nonumber &\times&\left.\int_r^R{\frac{\mathrm{d}x}{x}~\frac{v^2_c(x)}{v^2_R}\,\left[\frac{k_R}{k(x)}\right]^{3/5}\, 
\left[\frac{1 +\delta_R}{1 + \delta(x)}\right]^{3/5}}\right\}~.
\end{eqnarray}
Again, $n(r)$ is linked to $T(r)$ by $n(r) = [k_B T(r)/k(r)]^{3/2}$. In our
numerical computations we adopt the functional shape
\begin{equation}
\delta(r) = \delta_R\,e^{-(R-r)^2/\ell^2}
\end{equation}
which decays on the scale $\ell$ inward of a round maximum. The runs $\delta(r)$
we adopt are consistent with those indicated by numerical simulations (Lau
et al. 2009; Vazza et al. 2011). A power law has been instead used to describe
the radial distribution of the fraction $p_{\rm nth}/p_{\rm tot}$ by Morandi et al.
(2012; see also Shaw et al. 2010) in their 3-D structure reconstruction of
Abell 1835. They performed a triaxial joint analysis using X-rays, strong
lensing (SL) and SZ data available to infer the gas entropy and the
nonthermal pressure profiles out to $r_{200}$.

\begin{figure}
\epsscale{1}\plotone{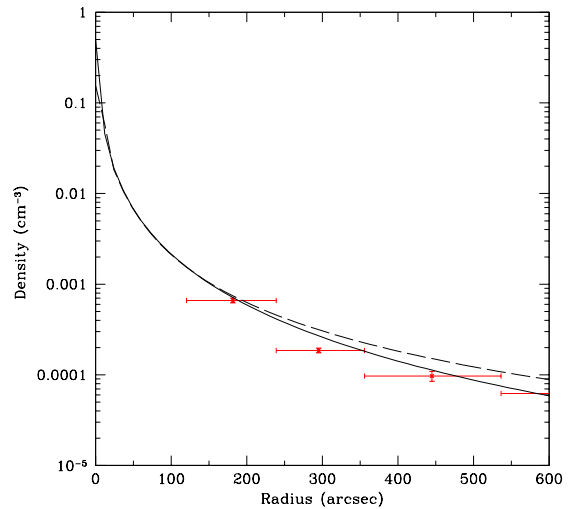}\caption{ICP density profile. Solid
line is the electron density profile obtained by the SM fit to the surface
brightness profile observed by \textsl{Chandra} in Abell 1835 (see Fig.~1); dashed
line is the fit with a double-$\beta$ model (Cavaliere \& Fusco-Femiano 1976)
of the deprojected density derived by Li et al. (2012) from the \textsl{Chandra}
results. The red points are the \textsl{Suzaku} results (Ichikawa et al. 2013).}
\end{figure}

\begin{figure*}
\begin{center}
\epsscale{1}\plottwo{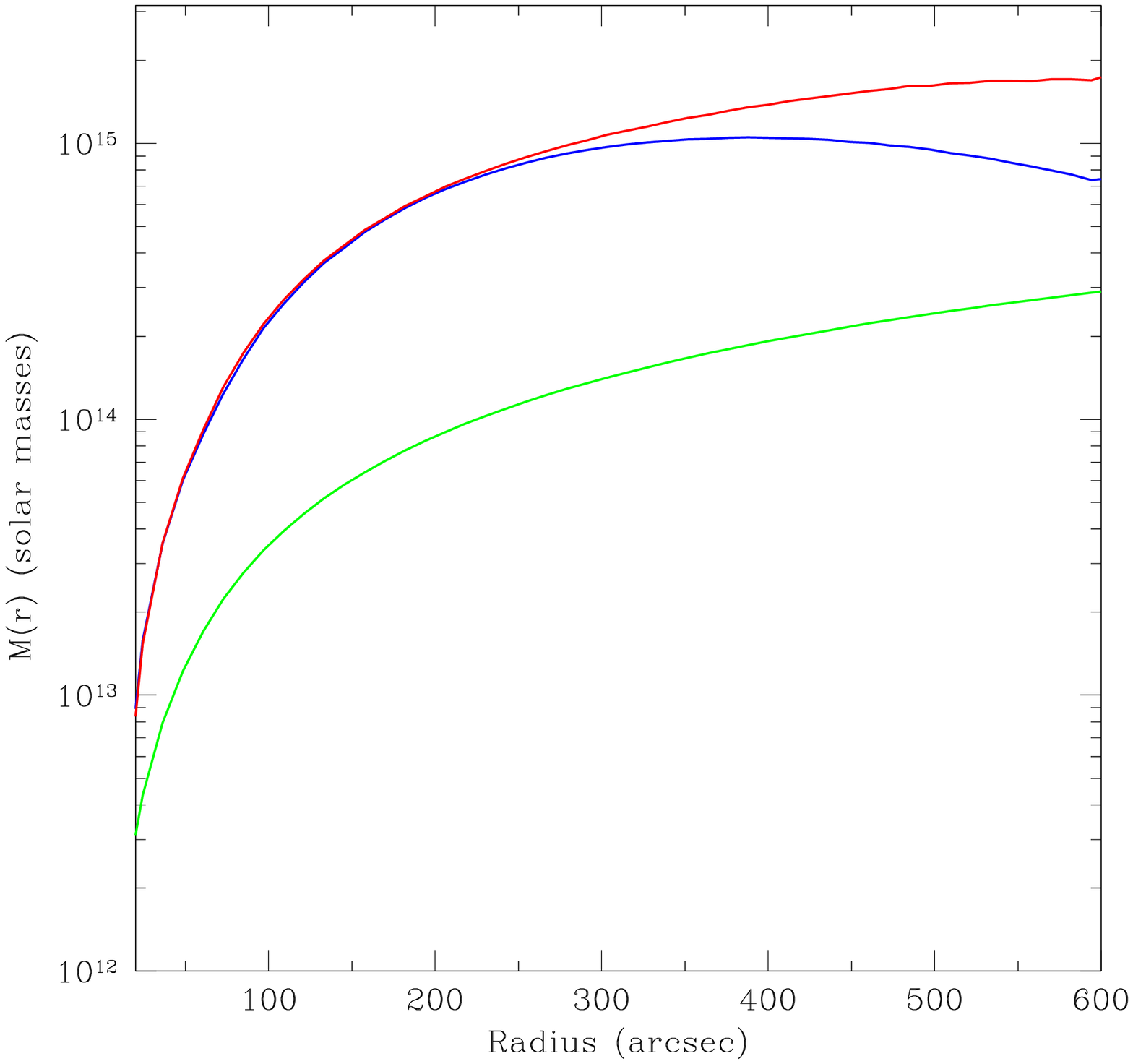}{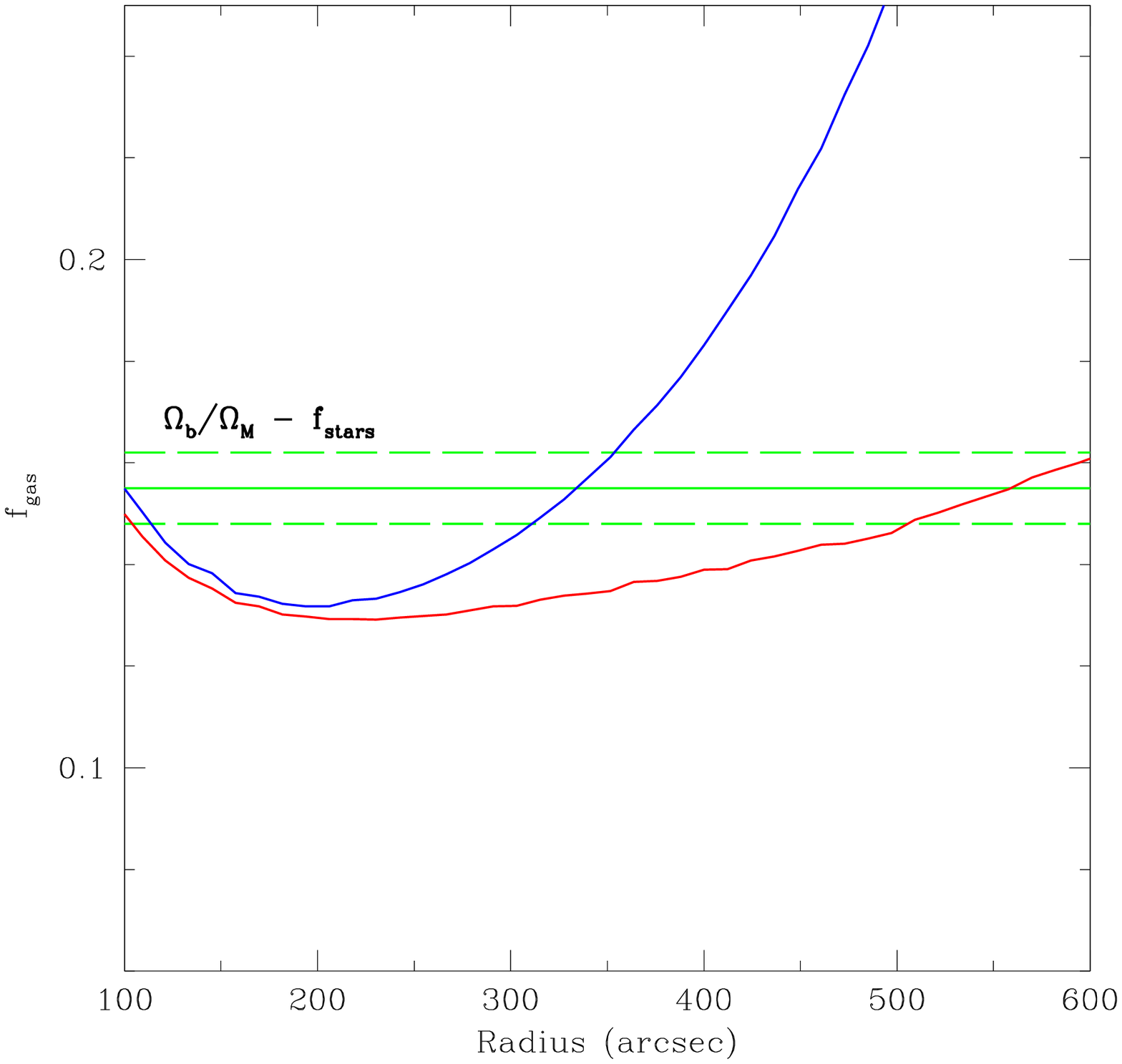}\caption{Left panel: Total cluster mass
and ICP mass for Abell 1835 derived from the SM analysis. Blue line is the
total mass obtained with $\delta(r) = 0$ (see Eq.~4); red line is the total
mass derived with $\delta_R$ = 1.4 and $\ell$ = 0.5 (see Eqs.~3 and 4); green
line is the gas mass derived from the gas density of Fig.~2 (solid line,
central value $n_{e,0}$ = 0.49 cm$^{-3}$). Right panel: Gas mass fraction
derived from the above mass profiles; blue line is with $\delta (r)$ = 0; red
line is with the above values of $\delta_R$ and $\ell$; green lines are the
difference of the cosmic baryon fraction and the fraction of baryons in stars
and galaxies, $\Omega_b/\Omega_M -f_{\rm stars} = 0.155 \pm 0.007$ (Komatsu et al.
2011; Gonzalez et al. 2007).}
\end{center}
\end{figure*}

\section{SuperModel Analysis of Abell 1835}

The SM analysis of Abell 1835 observed by \textsl{Chandra} (Schmidt et al. 2001;
Bonamente et al. 2013) begins assuming that the total pressure for the HE is
given only by the thermal ICP pressure (see Eq.~1). Fig.~1 shows the fit to
the projected temperature profile (blue line) assuming a deviation of the
entropy from the profile $k \sim r^a$ at $r > r_b$; this because a power law
increase is inconsistent with the \textsl{Chandra} data (see green line). From the
surface brightness distribution (see Fig.~1) we derive the ICP density
profile of Fig.~2 slightly different from the deprojected electron density
profile obtained by Li et al. (2012) from \textsl{Chandra} observations and in
agreement with the profile derived at $r \gtrsim 180^{\prime\prime}$ by the
\textsl{Suzaku} observations (Ichikawa et al. 2013). As shown in Fig.~5 the gas
density profile gives a central SZ effect value absolutely consistent with
the observations (Reese et al. 2002), at variance with the gas density profile
derived by Li et al. (2012). Moreover, our SZ effect profile
reproduces fairly well the profile observed by Bolocam at $r \gtrsim
30^{\prime\prime}$ (Sayers et al. 2011). The central gas density is 0.49
$\pm$ 0.03 cm$^{-3}$ while at the virial boundary ($R$ = 2.4 Mpc or 606.4
arcsec) is (5.73 $\pm$ 0.37)$\times$ 10$^{-5}$ cm$^{-3}$. This last value is
about a factor 2 greater than the gas density reported by Morandi et al.
(2012) at the virial radius. In accord with Bonamente et al. (2013), who have
analyzed the \textsl{Chandra} data with the fitting formulae of Vikhlinin et al.
(2006), the steep temperature profile causes a decreasing total matter at $r
\gtrsim$ 400$^{\prime \prime}$ and a consequent gas mass fraction consistent
with the cosmic value at $r = r_{500}$ ($\approx$ 327$^{\prime \prime}$), but
absolutely inconsistent at larger radii (see Fig.~3). This $M_{\rm tot}$ profile
provides evidence that beyond $r_{500}$ the HE is not supported only by
thermal pressure, as suggested by several theoretical studies (e.g., Lau
et al. 2009).

In Sect. 2 we have shown that the SM formalism has the ability to
straightforwardly include in the equilibrium a nonthermal pressure to yield
the total pressure $p_{\rm tot} = p_{\rm th}(1 + \delta)$ where the pressure $p_{\rm nth}
= p_{\rm th}\,\delta$ can be physically characterized in terms of a normalization
provided by the infall kinetic energy seeping through the virial shocks to
drive turbulence, and of a dissipative decay scale (see Eqs.~2 and 3). The
inclusion of a nonthermal component leads to an increasing total mass also in
the more peripheral regions of Abell 1835 (see Fig.~3, red line). We
determine the quantities $\delta_R$  and $\ell$ (see Eq.~3) imposing that the
baryon mass fraction equals the cosmic value at the virial radius (red line
in Fig.~3), and that the mass profile is smooth in the outskirts. These
values yield the pressure profiles $p_{\rm th}$, $p_{\rm nth}$ and $p_{\rm tot}$ shown in
Fig.~5 ($\delta_R$ = 1.4, $\ell$ = 0.5$R$). The thermal pressure is about 40$\%$
of the total pressure at the virial radius helped by turbulent motions in
sustaining the equilibrium, while it predominates at the center. The
nonthermal pressure starts to become significant at $r \gtrsim
400^{\prime\prime}$ where our analysis with $\delta = 0$, in accord with
Bonamente et al. (2013), reports a decreasing mass profile.

\begin{figure*}
\begin{center}
\epsscale{1}\plottwo{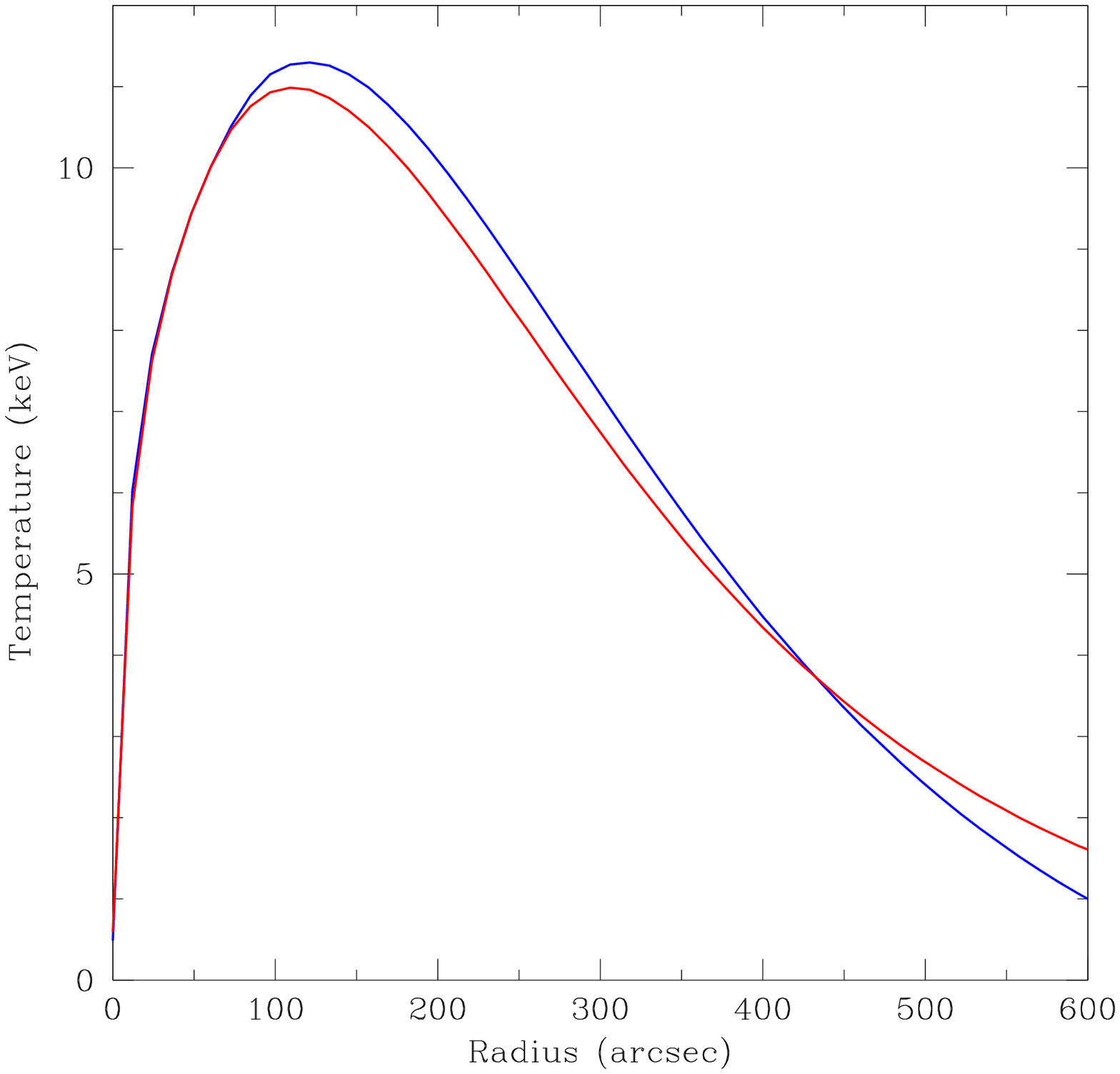}{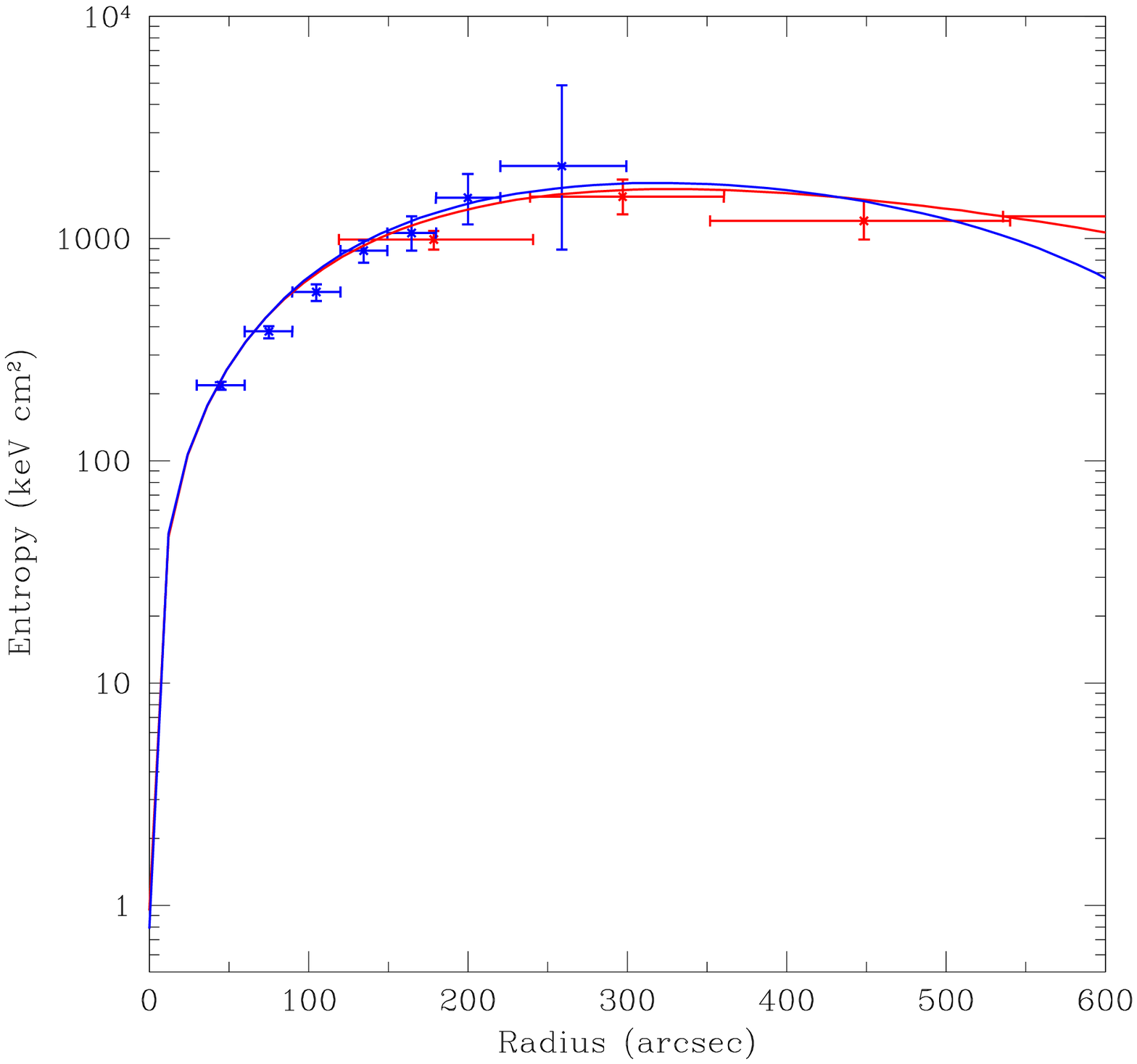}\caption{Left
panel: Radial temperature profile. Blue line is the radial temperature
obtained by the SM fit with $\delta(r) = 0$ to the projected profile observed
by \textsl{Chandra} in Abell 1835 (see Fig.~1); red line is from the SM fit with
$\delta_R$ = 1.4 and $\ell$ = 0.5. Right panel: SM entropy profile of Abell
1835. Blue points are reported by \textsl{XMM-Newton} (Zhang et al. 2007); red points are
from \textsl{Suzaku} (Ichikawa et al. 2013). Blue line is with
$\delta(r)$ = 0; red line is with $\delta_R$ = 1.4 and $\ell$ = 0.5.}
\end{center}
\end{figure*}

In presence of a nonthermal pressure the traditional equation to estimate the
total mass $M(r)$ within $r$ is modified as
\begin{eqnarray}
\nonumber M(r) &=& - \frac{k_B T(r)\,[1 +\delta(r)]\, r^2 }{\mu m_p G}\,\left\{
\frac{1}{n_e(r)}\frac{{\rm d} n_e(r)}{{\rm d} r}+\right.\\
&\nonumber&\\
\nonumber&+&\left.\frac{1}{T(r)[1+\delta(r)]}\frac{{\rm d} T(r)[1 + \delta(r)]}{{\rm d}r}\right\}=\\
&&\\
\nonumber&=& - \frac{k_B T(r)[1 +\delta(r)]\, r^2}{\mu m_p G}\left[\frac{1}{n_e(r)}
\frac{{\rm d} n_e(r)}{{\rm d} r}+\right.\\ 
&\nonumber&\\
\nonumber&+& \left.\frac{1}{T(r)}\frac{{\rm d} T(r)}{{\rm d} r} + 
\frac{\delta(r)}{1 + \delta(r)} \frac{2}{\ell^2}(R - r)\right]~,
\end{eqnarray}
where $k_B$ is the Boltzmann constant, $\mu$ is the mean molecular weight,
$m_p$ is the proton mass, $G$ is the gravitational constant. The mass of the
hot gas is
\begin{equation}
M_{\rm gas} = 4\pi\,\mu_e\, m_p\,\int{{\rm d}r~r^2\, n_e(r)}
\end{equation}
where $\mu_e$ is the mean molecular weight of the electrons.

The fit to the \textsl{Chandra} projected temperature profile with the the ratio
$\delta = p_{\rm nth}/p_{\rm th} > 0$ is only slightly different from the fit with
only the thermal pressure to sustain the HE. This difference is completely
negligible in the fit to the brightness profile due to its weak dependence on
the temperature (see Fig.~1). From these fits we extract values (with their
1-$\sigma$ uncertainty) of the parameters $k_c \approx 5 \pm 2$ keV cm$^2$,
$a \approx 1.29_{-0.48}$, and $k_R \approx 1040 \pm 520$ keV cm$^2$
specifying the entropy pattern for $r \leqslant r_b$; for $r > r_b$ the
entropy decline starts at $r_b\approx 0.11^{+0.16} R$ ($\approx 260^{+380}$
kpc), with a gradient $a^{\prime} \approx 0.47_{-0.33}$. Fig.~4 shows the 3-D
temperature and entropy profiles of the ICP when a nonthermal pressure
component is included in the HE equation. Our entropy profile for $\delta >
0$ is consistent with the observed entropy values derived by \textsl{XMM-Newton} (Zhang
et al. 2007) and \textsl{Suzaku} (Ichikawa et al. 2013) observations. Our value of
$r_b$ between $\approx$ (260-640) kpc derived by \textsl{Chandra} observations is
consistent with the radius in the interval $\approx$ (470-950) kpc where the
\textsl{Suzaku} entropy profile starts to decline downward (see Fig.~7 in Ichikawa
et al. 2013). We note also that the SM entropy profile (red line) is
sufficiently flat to satisfy the Schwarzschild criterion discussed by
Bonamente et al. (2013) for the convective instability. Moreover, an
increasing entropy profile that deviates from a power law is within the
uncertainty of the slope $a^{\prime}$.

\begin{figure*}
\begin{center}
\epsscale{1}\plottwo{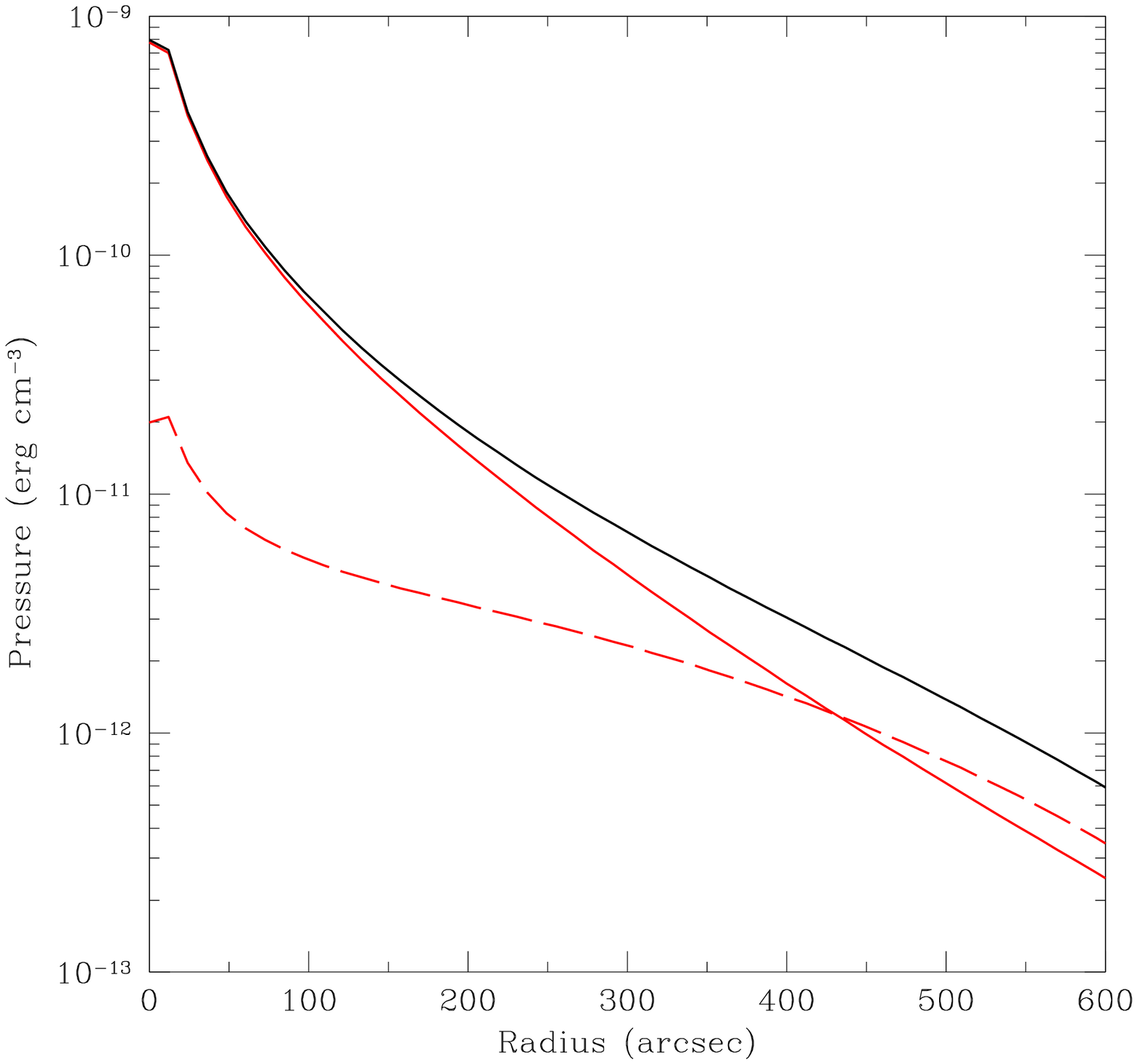}{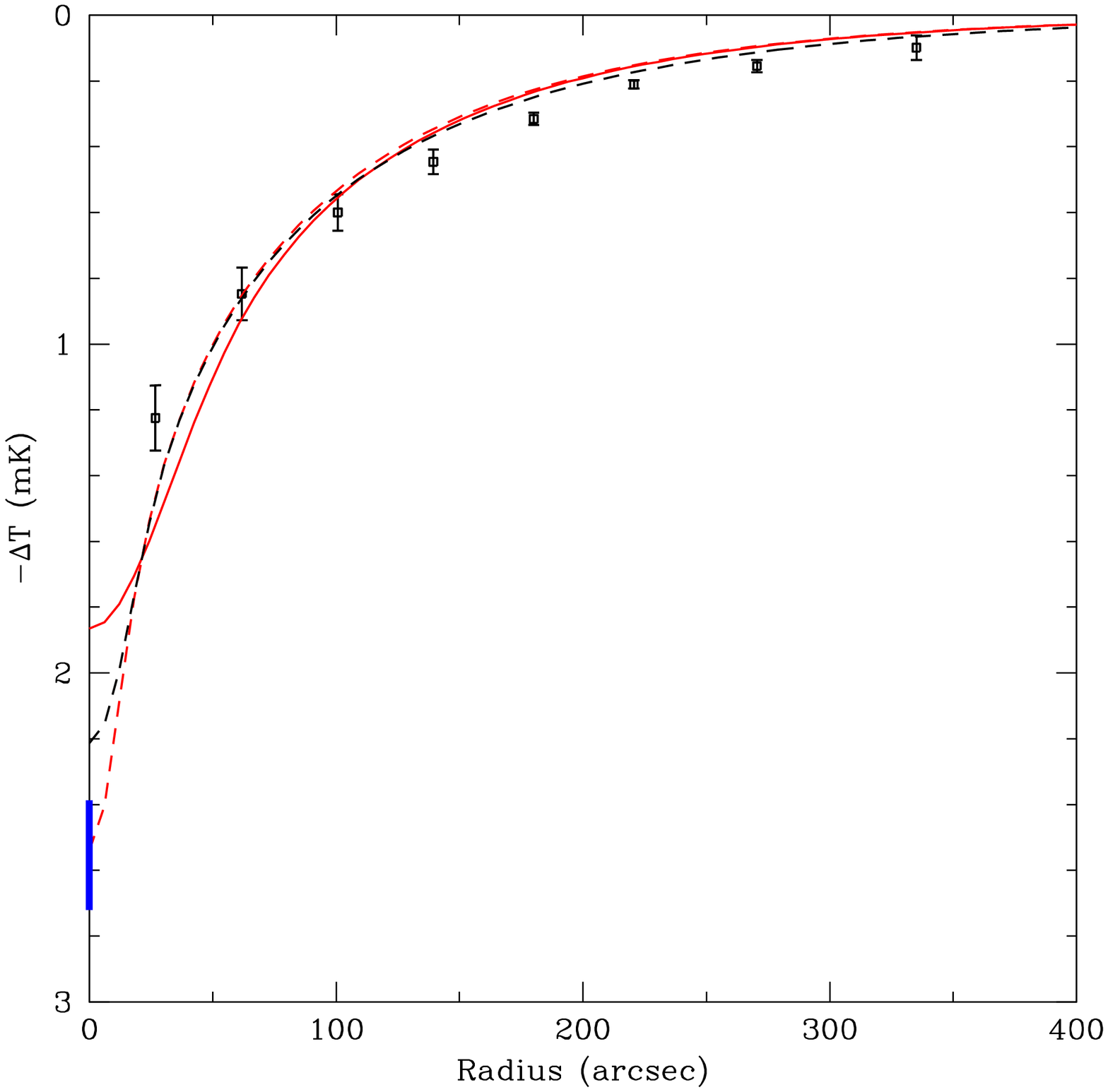}\caption{Left panel: Pressure profiles.
Red line is the thermal pressure; dashed line is the nonthermal pressure;
black line is the total pressure ($p_{\rm tot} = p_{\rm th} + p_{\rm nth}$). Right
panel: SZ effect in Abell 1835. Dashed red line is the SZ effect profile
obtained with the ICP density profile of Fig.~2 (solid line) and temperature
profile of Fig.~4 (red line, $\delta > 0$); dashed black line is the SZ
effect profile obtained with the gas density profile (dashed line of Fig.~2)
derived by Li et al. (2012) and temperature profile of Fig.~4 (red line).
These two profiles are compared with the central SZ effect value
($2.502^{+0.150}_{-0.175}$ mK, blue point) obtained by the \textsl{OVRO/BIMA}
interferometers with resolution $18^{\prime\prime}$ (Reese et al. 2002). Red
line is the SZ effect profile obtained with the ICP density profile of Fig.~2
(solid line) and temperature profile of Fig.~4 (red line, $\delta > 0$) to
compare with the black points observed by \textsl{Bolocam} at
$58^{\prime\prime}$ resolution (Sayers et al. 2011). All these profiles and
the data have been scaled to a frequency dependence of $-2$ of the thermal SZ
effect.}
\end{center}
\end{figure*}

\section{Discussion and Conclusions}

Only recently the use of the \textsl{Suzaku} observations and of the SZ effect
profiles have allowed to obtain some first insights on the thermodynamic
properties of the cluster outskirts. This avoids resorting to extrapolations
of the information available at $r \lesssim r_{500}$ to estimate the ICP and
total masses going toward the virial boundary. The \textsl{Suzaku} and \textsl{Chandra}
observations of several relaxed clusters have highlighted steep temperature
profiles and entropy profiles that deviate from the expected power law
increase (e.g., Walker et al. 2012). However, a recent combined analysis of SZ
and X-ray data seems not to indicate the entropy flattening in relaxed
clusters (Eckert et al. 2013). As already reported in Lapi et al. (2010) for a
number of clusters, here we confirm for Abell 1835 that the observed steep
temperature profile measured with \textsl{Chandra} can be fitted by our SM only
imposing a deviation of the entropy profile from a power law increase at $r >
r_b$. Also the recent \textsl{Suzaku} observations (Ichikawa et al. 2013) report an
entropy flattening; a similar behaviour is found in the combined X-rays, SL
and SZ data analysis of Morandi et al. (2012). We highlight that the goodness
of our gas density and temperature profiles obtained by the SM fits to the
\textsl{Chandra} X-ray observables is widely tested. The derived entropy profile is
in agreement with the entropy values reported by \textsl{XMM-Newton} and \textsl{Suzaku} results
(see Fig.~4), and the SZ effect profile is consistent with the
observations (see Fig.~5).

In the Perseus cluster, observed by \textsl{Suzaku} in the outskirts, the gas mass
fraction exceeds at the virial boundary the cosmic baryon value measured by
the CMB (Simionescu et al. 2011). The authors suggest that the most plausible
explanation for this apparent baryon excess toward the cluster periphery is
gas clumping. According to this interpretation the electron density is
overestimated affecting gas mass fraction, entropy, and pressure profiles.
The observed electron density must reach a value of up to $\sim$ 4 of the
true density at the virial radius to have $f_{\rm gas}$ consistent with the
cosmic value. However, as reported by Walker et al. (2012) the gas clumping
derived by Nagai \& Lau (2011) appears insufficient to match observations and
is expected to be most significant at $r \gtrsim r_{200}$ while the observed
entropy profiles start to flatten around 0.5$r_{200}$ (see also Ichikawa
et al. 2013). However, a recent paper by Walker et al. (2013) attributes to the
gas clumping the major responsibility  of the entropy flattening observed in
several clusters. But, for Abell 1835 the observed density needs to be
overestimated by a large factor $\sim$ 7 to make the entropy profile agree
with a power law increase in the outskirts. Besides, the too low measured
temperatures compared to predictions seem to be responsible for the entropy
flattening in this cluster.

The observed sharp drops in temperature imply decreasing mass profiles in the
outskirts of some relaxed galaxy clusters (e.g., Kawaharada et al. 2010;
Bonamente et al. 2013; Ichikawa et al. 2013). This unphysical situation may be
interpreted in terms of an ICP far from the HE. However, simulations show
that clusters are subject to an intense activity from the surrounding cluster
environment. Continued infall of gas onto clusters along filaments, violent
mergers of groups and sub-clusters, and supersonic motions of galaxies
through the ICP may induce turbulence that gives rise to a nonthermal
pressure (Lau et al. 2009; Burns et al. 2010; Vazza et al. 2011). The weakening
of the accretion shocks not only lowers the entropy production but also
increases the amount of bulk energy to drive turbulence into the outskirts
(Cavaliere et al. 2011). The result is that in addition to the thermal
pressure a nonthermal component may sustain the HE to obtain an increasing
mass profile and therefore a more accurate determination of the baryon gas
fraction.

We test this possibility in Abell 1835, observed by \textsl{Chandra} out to a radial
distance of $\sim$ 2.4 Mpc, exploiting the SM formalism; the latter is able
to include a nonthermal component (see Eq.~2), at variance with the fitting
formulae used in the analysis of the cluster X-ray observables by Bonamente
et al. (2013) and Landry et al. (2012). To determine the level and distribution
of the nonthermal pressure that in addition to the thermal pressure sustains
the HE we have imposed that the gas baryon fraction equals the observed
cosmic value at the virial radius $R$. Our constraint is supported by the
\textsl{Suzaku} observations that report a gas mass fraction, defined by the lensing
total mass, that at $R$ agrees with the cosmic baryon fraction. Also the
combined analysis of Eckert et al. (2013) reports that at $r_{200}$ the gas
fraction converges for relaxed clusters to the expected value.

The thermal and nonthermal pressure profiles of Fig.~5 define the total
pressure distribution that guarantees HE everywhere as evidenced by the
increasing profile of the cluster mass (see Eq.~2 and Fig.~3). The goodness
of the SM analysis is confirmed by the comparison between our total mass
values at $r_{500}$ and $R$ with the weak lensing cluster mass measured by
Clowe \& Schneider (2002) and Hoekstra et al. (2012). In particular, the last
authors report $M^{NFW}_{\rm vir} = 1.89^{+0.38}_{-0.35}\times 10^{15} M_{\odot}$
consistent with our value of $\sim 1.75\times 10^{15} M_{\odot}$ obtained
with $\delta > 0$ and inconsistent with the value of $\sim 7.50\times 10^{14}
M_{\odot}$ derived when the HE is supported only by the thermal pressure
($\delta = 0$).

For Abell 1835 we obtain a nonthermal pressure contribution at the virial
radius around 60$\%$ of the total pressure and $\ell\sim 0.5\, R$ in agreement
with the simulations of Burns et al. (2010) that report $p_{\rm nth}/p_{\rm tot}
\approx (60-65)\%$ for a sample of clusters. The ratio between the mass
estimated including turbulence in the SM and the mass estimated without
turbulence $M_{\rm turb}/M_{\rm noturb}$ is $\sim$ 2.4. Giles et al. (2012) found that
X-ray hydrostatic masses for relaxed clusters are underestimated by a factor
1.21$\pm$0.23 when compared to the weak-lensing masses. The level of the
nonthermal pressure at the virial radius and the ratio $M_{\rm turb}/M_{\rm noturb}$
are strictly related to the ICP temperature run that is the main responsible
for the mass profile. The above values are justified by the uncommon drop of
a factor $\sim$ 10 from the peak temperature to the value at the virial
radius reported by \textsl{Chandra} in Abell 1835 (see Fig.~1). A lower value ($\sim
5$) is reported by \textsl{Suzaku} (Ichikawa et al. 2013). For a drop factor
of $\sim$ 2.5, more similar to those reported by \textsl{Suzaku} observations in
other clusters, we obtain that the nonthermal pressure at the virial radius
decreases to $\sim 35\%$ of the total pressure and $M_{\rm turb}/M_{\rm noturb}$
lowers to $\sim$ 1.31 consistent with the average value derived by Giles
et al. (2012). For this smoother decline of the temperature profile the
nonthermal pressure contribution to the total support is consistent with that
derived by simulations for relaxed clusters (see Lau et al. 2009; Vazza et al.
2011). These simulations show a radial increase of $\delta$ similar to that
described by Eq.~3 and a nonthermal pressure contribution to the total
pressure of (30-40)$\%$ at the virial boundary. Greater values are obtained
in the simulations of some relaxed clusters. A lower level of about 20\% has
been derived by the analysis of Morandi et al. (2012). A value that is also
lower than the predictions from numerical simulations. This discrepancy may
be due to their use of X-ray data limited at $r_{500}$ where the steepening
of the temperature profile observed by \textsl{Chandra} and \textsl{Suzaku} is not yet
evident. A further cause is to consider spherical averaging of ellipsoidal
galaxy clusters in the context of X-ray observables. However, the mean biases
in observables are not greater than few percent within $r_{500}$ (Buote \&
Humphrey 2012), although higher values are likely going toward the virial
radius.

Mahdavi et al. (2013) found relaxed clusters consistent with no bias when
hydrostatic and weak lensing masses are compared at $r_{500}$. But, we
believe that the increasing radial profile of $p_{\rm nth}$ reported by the
simulations may give hydrostatic masses that bias low at the virial radius.
This is supported by the differences between $M_{\rm noturb}$ and $M_{\rm turb}$ at
$r_{500}$ and $R$ in the relaxed Abell 1835 (see Fig.~3). This difference is
negligible at $r_{500}$ and much evident at $R$.

In summary, we have shown how the analysis of the X-ray observables allows to
derive a total mass profile consistent with the weak lensing measurements,
and to trace the thermal and nonthermal pressure profiles. This can be
obtained on introducing in the HE equation a nonthermal pressure support as
allowed by our SuperModel. In particular, we have reconstructed from \textsl{Chandra}
X-ray observations the gas and total mass profiles of Abell 1835. The values
of $\delta_R$ and $\ell$ that defines the nonthermal pressure component have
been obtained by the condition that $f_{\rm gas}$ equals $\Omega_b/\Omega_M$ -
$f_{\rm stars}$ at the virial radius. We have also shown that the level of
turbulence $\delta_R$ depends on the observed ICP temperature profile. A
steep drop in $T$ implies a decreasing mass profile and therefore a high
level of turbulence is required to obtain an increasing cluster mass profile
that satisfies the cosmic gas mass fraction at the virial boundary. A
lower level is necessary for a smoother decline of the temperature. This is
consistent with the weakening of the accretion shocks that leads to a
reduction of the thermal energy to feed the ICP entropy and to an increase of
the bulk energy to drive turbulence in the cluster outskirts. As discussed in
Sect. 2, the weakening degree of the accretion shocks may depend on the
cluster environment and this seems to be confirmed by the significant
azimuthal variations of the electron density, temperature and entropy
reported by \textsl{Suzaku} (Ichikawa et al. 2013). Using the SDSS photometric data
for Abell 1835 and Abell 1689, the authors found that the hot regions are
associated with a filamentary structure, while the cold regions contact
low-density regions outside the clusters. Finally, the increasing $f_{\rm gas}$
profile at $r \gtrsim 0.3R$ reported in Fig.~3 confirms the conclusion of
Rasheed et al. (2010) that the baryons are not missing. They are simply
located in the most peripheral regions of the clusters likely for the heating
processes (such as shocks-heating of the gas, supernovae and AGN feedback)
that cause the ICP to expand or hinder its inflow.

\begin{acknowledgements}
We thank our referee for constructive comments. We are grateful to
Massimiliano Bonamente for the submission of data on the \textsl{Chandra} brightness
distribution and to Mauro Sereno for clarifying discussions. Work supported
by INAF and MIUR. A.L. thanks SISSA for warm hospitality.
\end{acknowledgements}

\end{document}